# RR LYRAE VARIABLE STARS: PULSATIONAL CONSTRAINTS RELEVANT TO THE OOSTERHOFF CONTROVERSY

Giuseppe Bono

Osservatorio Astronomico di Trieste, Via G.B. Tiepolo 11, 34131 Trieste, Italy;
bono@oat.ts.astro.it

Filippina Caputo

Istituto di Astrofisica Spaziale, Via E. Fermi 21, 00044 Frascati, Italy

Vittorio Castellani

Dipartimento di Fisica, Univ. di Pisa, Piazza Torricelli 2, 56100 Pisa, Italy;
vittorio@astr1pi.difi.unipi.it

and

Marcella Marconi

Dipartimento di Fisica, Univ. di Pisa, Piazza Torricelli 2, 56100 Pisa, Italy;
marcella@astr1pi.difi.unipi.it

# ABSTRACT

A solution to the old Oosterhoff controversy is proposed on the basis of a new theoretical pulsational scenario concerning RR Lyrae cluster variables (Bono and coworkers).

We show that the observed constancy of the lowest pulsation period in both Oosterhoff type I (OoI) and Oosterhoff type II (OoII) prototypes (M3, M15) can be easily reproduced only by assuming the canonical evolutionary horizontal-branch luminosity levels of these Galactic globular clusters and therefore by rejecting the Sandage period shift effect (SPSE).

*Subject headings:* globular clusters: individual (M3, M15) – stars: evolution – stars: horizontal branch – stars: oscillations – stars: varables: other

# 1. Introduction

More than fifty years ago Oosterhoff (1939, 1944) brought to light a curious occurrence which opened a debate which has continued until the present years. In brief, let us recall that globular stellar clusters belonging to the galactic halo show the occurrence of radially pulsating stars, classified as RR Lyrae variables. Almost one century ago Bailey (1899, 1902a, b) recognized the occurrence in the RR Lyrae group of different types of pulsators, which we now understand in terms of stars pulsating either in the fundamental (RR*ab*) mode or in the first overtone mode (RR*c*, Schwarzschild 1941). According to Oosterhoff, globular clusters which present RR Lyrae variables can be separated into two different classes, the so called Oosterhoff type I (Oo I) and type II (Oo II), according to the mean periods of type *ab* pulsators (Fig. 1). Clusters characterized by lower metal abundance (Z), such as $M15$, show mean periods larger (Oo II group) than the periods of less metal-poor clusters, such as $M3$, belonging to Oo I group.

A vast production of literature has been devoted to the mysterious origin of the so-called Oosterhoff dichotomy. However, the question has become even hotter since the 1980s, when Sandage (1981, 1982) suggested that such a difference between the mean periods was the consequence of a similar difference in periods shown by single pulsators at a fixed effective temperature. To understand the disruptive consequence of such a suggestion, we may recall the well-established relation connecting the period of a fundamental pulsator to the stellar evolutionary parameters:

$$\log P = 11.497 + 0.84 \log L - 0.68 \log M - 3.48 \log T_e,$$

where both the mass ($M$) and the luminosity ($L$) are in solar units whereas the effective temperature ($T_e$) is in Kelvins. According to Sandage's suggestion, when comparing pulsators with the same temperature in M3 and M15 clusters, the period of M15 pulsators turns out to be larger by $d \log P = 0.055$, which is well outside the value provided for these stars by current evolutionary theories. In short, let us stress that, if the scenario outlined by Sandage is right, the evolutionary predictions about the luminosity of these stars, (the horizontal branch [HB] stars) should be wrong. That would be a rather dramatic conclusion, since the evolutionary properties of these stars are at the basis of current constraints about relevant cosmological parameters such as the age of the universe and the amount of cosmological helium.

Thus in literature much debate is devoted to discussion of the observational basis of Sandage's suggestions and, in particular, the assumptions adopted for deriving the temperature scale (Caputo 1988; Fernley 1989; Caputo & De Santis 1992; Carney, Storm,

& Jones 1992; Carney *et al.* 1992). In this Letter we present a new piece of theoretical evidence which can help to solve the question in the direction of a correct evolutionary approach.

## 2. Discussion

As a starting point, let us recall that, according to a well established procedure, the *troublesome* occurrence of two types of pulsators can be easily overcome by transforming the periods of first-overtone pulsators into the corresponding periods of the fundamental mode (van Albada & Baker 1973). On this basis, a very interesting occurrence in the distribution of *fundamentalized* periods was found some years ago by Caputo, Castellani & Tornambè (1989). The period distribution of the two Oosterhoff group prototypes, $M3$ and $M15$, overlaps nicely at the lower periods (Figure 2). This implies that stars at the blue edge of the instability strip (i.e. at the maximum surface temperature for pulsation instability) present very similar periods. However, at that time this was presented only as a suggestion (a strong suggestion in our thought) in favor of the evolutionary scenario and against Sandage's suggestion. As a matter of fact, at that time we were lacking a firm evaluation of both the location of the instability strip in the H-R diagram and of the blue-edge dependence on surface He abundance and/or mass of the pulsators.

As a final step in a long-pursued program about the properties of pulsators based on nonlinear, nonlocal, and time-dependent convective models (Bono & Stellingwerf 1994; Bono, Caputo, & Stellingwerf 1994a,b; Bono *et al.* 1995a,b) we have now such evaluations, which can be used to throw light on Sandage's suggestions. According to these pulsational results, it has been found that for each given luminosity level the temperature of the blue boundary of the instability strip increases when the mass of the pulsator is increased, whereas no relevant dependence was found on the abundance of surface He. As a consequence, the fundamentalized period of a pulsator located to the right of the blue boundary can be safely described by the following relation (Fig. 3):

$$\log P = 1.03 \log L - 0.64 \log M - 2.26$$

This allows us to go straight to the question. Let us first check the *incipit* of stellar evolutionary theory. According to current estimates (Castellani, Chieffi & Pulone 1991) the stellar mass and the luminosity level for $M15$ ($Z = 0.0001$) are $M = 0.77$, $\log L = 1.70$ respectively, whereas for $M3$ ($Z = 0.0004$) these values are $M = 0.68$ and $\log L = 1.65$. Once these values are inserted in the previous relation, we find that the evolutionary expectation concerning the minimum period variation is of the order of $d \log P = 0.009$, that is, a quite negligible variation. Moreover, it is worth noting that this minimum value is located close to $\log P = -0.44$, again in excellent agreement with observations. Therefore, at least in this respect, the evolutionary theory is in agreement with observational data.

Now Sandage's hypothesis can be checked. On the basis of both the observed constancy of the smallest period and the suggested Sandage period-shift effect (SPSE), the following relations were derived

$$1.03 d\log L - 0.64 d\log M = 0$$

$$0.84 d\log L - 0.68 d\log M = 0.055$$

which present a *formal* solution only if the differences ($M15 - M3$) are of the order of $d\log L = -0.22$ and $d\log M = -0.35$ (!), well beyond any plausible limit. As an example, since pulsators masses in $M3$ can be safely assumed to lie in the range 0.6-0.8 $M_\odot$, this would require mass values lower than or of the order of 0.35 $M_\odot$ for the RR Lyrae variables belonging to M15. But this value cannot be accepted under any reasonable evolutionary assumption.

Indeed, an increased difference in luminosity between OoI and OoII clusters, as is usually accepted to account for the SPSE, does not go in the right direction, since it would imply a not negligible increase in the OoII minimum period, which would be in contrast to the observations. Even a reasonable increase in the mass of the OoII pulsators cannot be of any use. Only an increase in periods dominated by an unacceptable decrease in the pulsator masses, moderated by a corresponding decrease in luminosity, could account for the observed constancy of the minimum period.

Therefore the pulsational scenario can be saved only by rejecting the SPSE and also any attempt to move the luminosity of horizontal-branch stars beyond their well-established theoretical evaluation. At the same time the current cosmological constraints provided by this kind of old star are still retained.

In conclusion, the results presented in this paper show quite clearly how both canonical HB evolutionary tracks and nonlinear pulsation models at full amplitude match the observed properties of Galactic globular clusters. Finally, let us emphasize that the evolutionary masses adopted in the present investigation for cluster variables are in agreement with the recent results provided by Cox (1991) and Kovacs *et al.* (1992) which firmly remove the long-standing mass discrepancies between pulsation and evolution theories and between globular clusters of different Oosterhoff types.

## 3. Figure Captions

**Fig. 1.** - The distribution of RRab mean periods in 24 galactic globular clusters characterized by more than 10 RR Lyrae variables (Castellani & Quarta 1987), disclosing the Oosterhoff dichotomy.

**Fig. 2.** - The comparison between the period frequency histograms for RR Lyrae in $M3$ (dashed line) and $M15$ (solid line).

**Fig. 3.** - Fundamentalized period along the first overtone blue boundary versus luminosity at fixed chemical composition ($X = 0.76$, $Z = 0.0001$) and for two different values of the stellar mass.

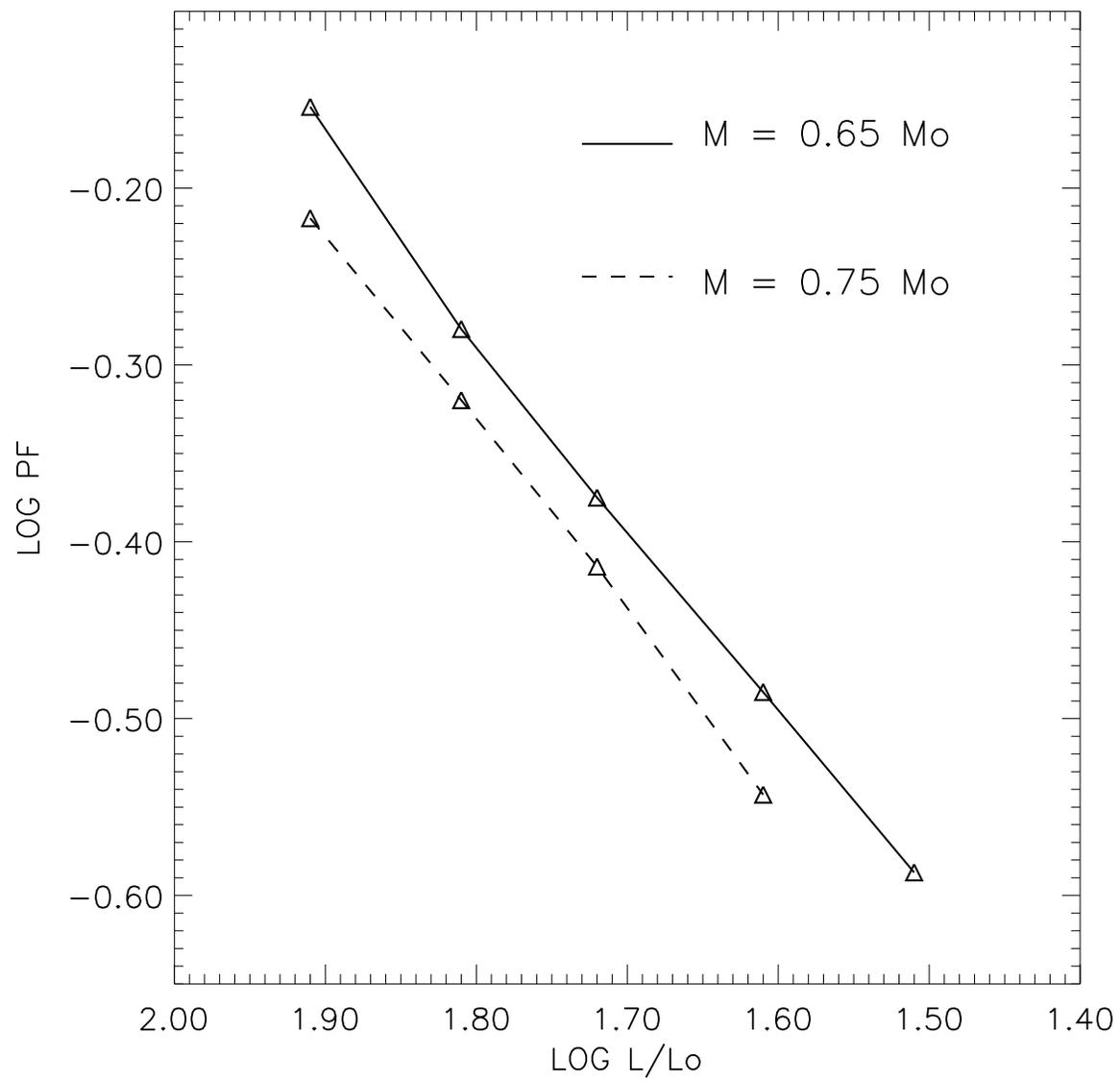

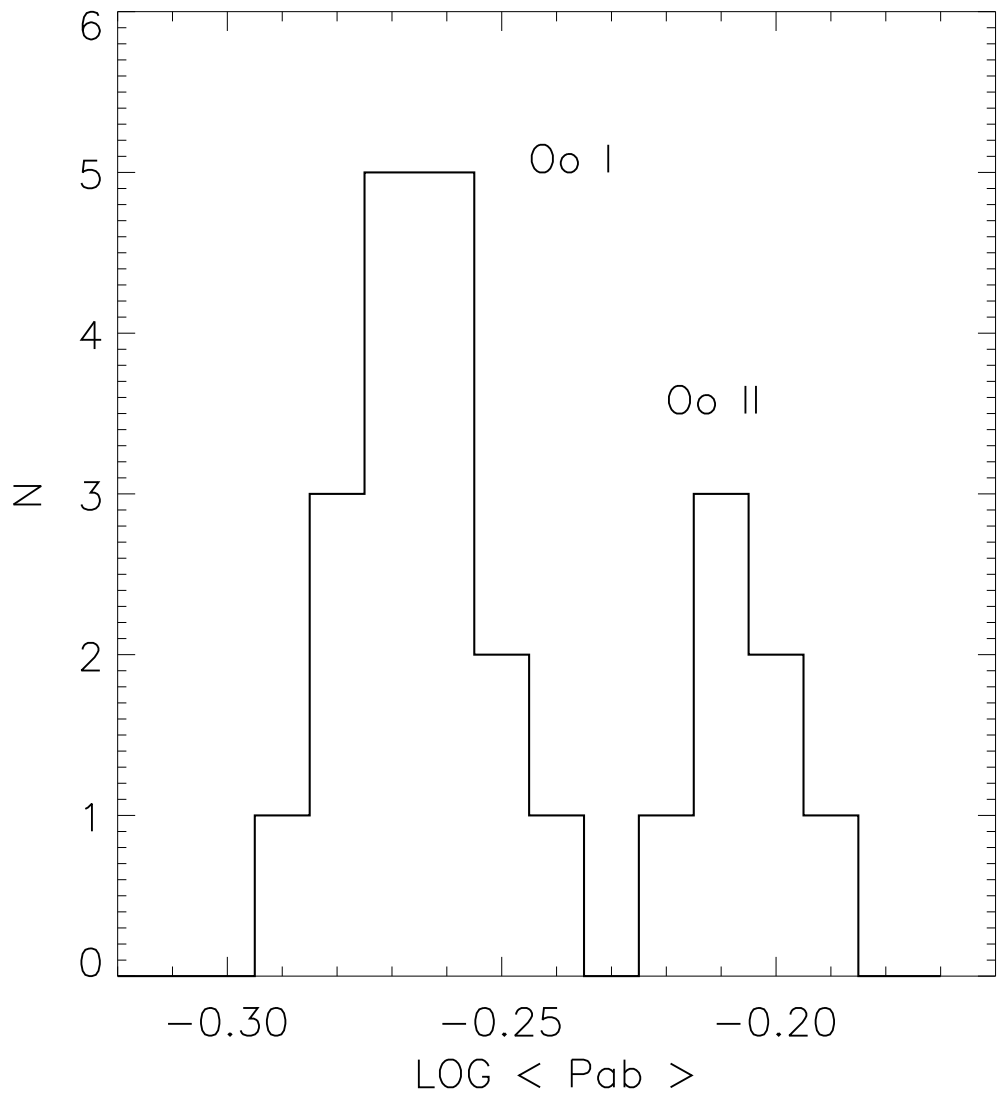